\documentclass[aps,prl,twocolumn,superscriptaddress]{revtex4}
\usepackage{epsf,graphicx}
\usepackage{amssymb}
\usepackage{amsmath}
\usepackage{bm}
\usepackage{color}
\usepackage{ulem}
\begin{document}
\title{Monte Carlo approach to quantum work in strongly correlated electron systems}
\author{Qian-Xi Zhao}
\affiliation{Department of Physics and Chongqing Key Laboratory for Strongly Coupled Physics, Chongqing University, Chongqing 401331, People ’s Republic of China}
\author{Jian-Jun Dong}
\email[Contact author: ]{dongjianjun@cqu.edu.cn}
\affiliation{Department of Physics and Chongqing Key Laboratory for Strongly Coupled Physics, Chongqing University, Chongqing 401331, People ’s Republic of China}
\author{Zi-Xiang Hu}
\email[Contact author: ]{zxhu@cqu.edu.cn}
\affiliation{Department of Physics and Chongqing Key Laboratory for Strongly Coupled Physics, Chongqing University, Chongqing 401331, People ’s Republic of China}
\date{\today}

\begin{abstract}
We develop a Monte Carlo framework to analyze the statistics of quantum work in correlated electron systems. Using the Ising-Kondo model in heavy fermions as a paradigmatic platform, we thoroughly illustrate the process of determining the moment generating function of quantum work under nonequilibrium conditions in detail. Based on this function, we systematically investigate essential statistical quantities, including the mean irreversible work density, the mean work density, variance, and the third central moment of quantum work across different quench processes. Our findings highlight distinct singularities in these quantities at the metal-insulator phase transition point at low temperatures. However, these singularities disappear, and the transition becomes a smooth crossover at high temperatures. This stark contrast underscores quantum work as an effective thermodynamic tool for identifying metal-insulator phase transitions. Our approach provides a promising new framework for investigating nonequilibrium quantum thermodynamics in strongly correlated electron systems.
\end{abstract}

\maketitle

Fueled by experimental advances and the discovery of fluctuation theorems, quantum thermodynamics has witnessed significant attention and advancement in recent years \cite{Binder2018,Deffner2019,Alicki2023PRD,Gatto2024PRA,WeiWu2024PRL}.
As an emerging research field, quantum thermodynamics is devoted to unraveling the intrinsic relationship between the principles of thermodynamics and their roots in quantum mechanics. It aims to build a new thermodynamic framework that goes beyond the conventional regime of validity of macroscopic thermodynamics \cite{Kosloff2013Entropy,Vinjanampathy2016ContempPhys,Millen2016NJP,Dann2023NJP}. The interdisciplinary character of this field has led to its extensive application in various domains, including black hole physics, quantum field theory, statistical mechanics, quantum information theory, and condensed matter physics \cite{Ortega2019PRL,Beyer2020PRR,Pourhassan2021JHEP,Gardas2016SciRep,Wei2018PRE,Zhou2021PRE,Erdamar2024PRR,Lacerda2023PRB,Lacerda2024PRE,Binder2023PRL,Gherardini2024PRXQuant,Davoudi2024PRL}.

A central object in quantum thermodynamics is the notion of work associated with the dynamics of an isolated quantum system \cite{Lobejko2017PRE,Francica2023PRB,Santini2023PRB,Kiely2023PRR,Pal2023PRB,Zawadzki2023PRA}. This is because the distribution of quantum work yields extensive insights into a non-equilibrium process, much like the way the partition function contains critical details about an equilibrium state. By means of the quantum Jarzynski equality and other fluctuation theorems, the equilibrium quantities, such as free energy difference, can be related to the quantities outside equilibrium, namely the averaged exponentiated work done during the process \cite{Jarzynski1997PRL,Jarzynski1997PRE}. As a result, quantum work can not only describe the equilibrium quantum phase transitions and the critical phenomena associated with the quantum criticality \cite{Silva2008PRL,Fei2020PRL,Wu2025PRL,Solfanelli2025PRL}, but also characterize the non-equilibrium dynamical quantum phase transitions \cite{Heyl2013PRL}. Furthermore, it has been suggested that quantum work is also related to the Loschmidt echo and information scrambling \cite{Campisi2017PRE,Chenu2018SciRep}, exhibiting the wide applicability of quantum work.

Recognizing the importance of quantum work, a variety of approaches have been proposed to calculate the work statistics, including the Feynman-Kac quantum equation \cite{FeiLiu2012PRER}, the path-integral approach \cite{Funo2018PRL,Dong2019PRB,Quan2020PRE}, the phase-space formulation \cite{FeiLiu2019PRE}, the group-theoretical approach \cite{Quan2019PRR}, the non-equilibrium Green's function method \cite{Quan2020PRL}, the random matrix theory \cite{Grabarits2022SciRep}, and the tensor-network approach (matrix-product-state method) \cite{Quan2022PRR,FengLiLin2024PRR}. However, these investigations concentrate on single-particle systems, non-interacting or weakly coupled many-body systems with perturbative driving protocols, one-dimensional quantum spin chains, and other models that can be exactly solved. The development of a systematic approach for computing the work statistics of strongly correlated electron systems remains an open challenge.

To address this issue, we introduce the famous Monte Carlo approach to calculate the moment generating function of quantum work. The Monte Carlo method, despite encountering the sign problem, has been extensively applied to many-body systems as an effective tool \cite{Gubernatis2016,Farkasovsky2010PRB,Kato2010PRL,Ishizuka2012PRL,Maska2020PRB,Dong2021PRBL,Dong2022PRBL,Qin2023PRB}. However, it has not been applied explicitly to quantum work so far. In this work, we illustrate this method in detail and show how to calculate the quantum work successfully by taking the Ising-Kondo model as a concrete example. For typical quench processes, we found that at low temperatures, the mean irreversible work density, the mean work density, the variance, and the third central moments of quantum work exhibit clear singularities at the metal-insulator phase transition point. While the singularities vanish at high temperatures due to a smooth crossover instead of a phase transition, this highlights the effectiveness of quantum work as a thermodynamic tool for identifying the metal-insulator phase transition. Our work initiates the exploration of quantum thermodynamics in strongly correlated electron systems based on Monte Carlo method.

We begin with a general definition of quantum work. For a closed system, the quantum work $W$ is defined by the difference of two consecutive energy measurements, that is, $W=E_{m}^{f}-E_{n}^{i}$. Here, $E_{n}^{i}$ and $E_{m}^{f}$ are the $n$th and $m$th eigenvalues of the initial and final Hamiltonian $H\left(  J\left(  t=0\right)  \right)  $, $H\left(  J\left(t=\tau \right)  \right)  $, respectively. In the following, we do not write the time dependence of the control parameter $J(t)$ explicitly and define $H(t)\equiv H(J(t))$ for compact notation. At nonzero temperature, the quantum work depends not only on the change protocol of the control parameter $J\left(  t\right) $, but also on the initial distribution of states, and therefore it is a stochastic variable with a probability distribution, which is given by
\begin{equation}
P\left(  W\right)  =\sum_{n,m}\delta(W-E_{m}^{f}+E_{n}^{i})P\left(  m^{f}\mid n^{i}\right)  P\left(  n^{i}\right)  , \label{wdf}
\end{equation}
where $P\left(  n^{i}\right)  =\exp\left(  -\beta E_{n}^{i}\right)  / Z\left(0\right) $ is the probability of sampling the $n$th eigenstate from the initial thermal equilibrium state at inverse temperature $\beta$ when making the initial energy measurement and $Z\left(  0\right) = \operatorname*{Tr}\left[  \operatorname{e}^{-\beta H\left(  0\right)  }\right]  $ is the corresponding partition function. $P\left(  m^{f}\mid n^{i}\right)  \equiv\left\vert \left\langle m^{f} \left\vert U\left(  \tau,0\right)  \right\vert n^{i}\right\rangle \right\vert^{2}$ is the transition probability between the initial states $\left\vert n^{i}\right\rangle $ and the final states $\left\vert m^{f}\right\rangle $ governed by the time-dependent evolution $ U\left(  \tau,0\right)  =\mathcal{T}\exp [  -\operatorname*{i}\int_{0}^{\tau}dtH\left(  t \right)  ]  $, where $\mathcal{T}$ is the time-ordering operator and we have set $\hbar=1$.

Generally, directly calculating the distribution function $P(W)$ presents challenges. Therefore, we focus on the moment generating function of the work distribution, which is defined as
\begin{align}
G\left(  R\right)   &  =\int_{-\infty}^{\infty}dW \operatorname{e}^{R W}  P\left(  W\right) \nonumber\\
&  =\frac{\operatorname*{Tr}\left[  U^{\dagger}\left(  \tau,0\right) \operatorname{e}^{R H\left(  \tau\right)  }U\left(  \tau,0\right)
\operatorname{e}^{-\left( \beta+R\right)  H\left(  0\right) }\right]  }{\operatorname*{Tr}\left[  \operatorname{e}^{-\beta H\left( 0\right)  }\right]  }. \label{gfgeneral}%
\end{align}
The moment generating function $G\left(  R\right)$ contains all the key information characterizing the dynamical process. Explicitly, the distribution function $P(W)$ is given by the inverse solution, and the moments of quantum work are related to the derivatives, $\left\langle W^{n}\right\rangle =\frac{d^{n}G\left(R\right)  }{dR^{n}}|_{R=0}$. The moment generating function is the quantity that we will calculate with the Monte Carlo method and the numerical results must satisfy the properties $G\left(  R\right)  >0$ and $G\left(  0\right)  =1 $.

We demonstrate how to compute the moment generating function $G\left( R \right)$ for a strongly correlated heavy fermion system utilizing the Monte Carlo method, specifically the Ising-Kondo lattice model on a square lattice with nearest-neighbor interaction \cite{Sikkema1996PRB,Zhong2019PRB,Zhong2021PRB,Zhou2024PRB},
\begin{equation}
H=-g\sum_{\langle ij\rangle\sigma}\left(  c_{i\sigma}^{\dagger}c_{j\sigma }+\text{H.c.}\right)  +J\sum_{j}s_{j}^{z}S_{j}^{z}, \label{Ising-Kondo}
\end{equation}
where $c_{j\sigma}^{\dagger},c_{j\sigma}$ are the creation and annihilation operators for conduction electrons, $s_{j}^{z}=\sum_{\alpha\beta}c_{j\alpha}^{\dagger}\frac{\sigma_{\alpha\beta }^{z}}{2}c_{j\beta}$ is the spin operator for conduction electron spin with $z$-component of Pauli matrixes $\sigma^z$, $S_{j}^{z}$ is the local spin of the $f$-electron, $g$ is the hopping coefficient between nearest-neighbor conduction electrons, and $J$ is the Kondo exchange coupling. In the following, we set $g=1$ and take $J(t)$ as the control parameter.

Noticing that the trace of the numerator of $G\left(  R\right)$ can be decomposed into contributions from the conduction electrons and the local spins of $f$-electrons,
\begin{equation}
\widetilde{G}\left(  R\right)  =\operatorname{Tr}_{c}\operatorname{Tr}_{S}\left[  U^{\dag}\left(  \tau,0\right)  \operatorname{e}^{RH\left(  \tau\right)}U\left(  \tau,0\right)  \operatorname{e}^{-\left(  \beta +R\right)  H\left(0\right)  }\right].
\end{equation}
In this system, the $f$-electron spin moment at each site is conservative since $\left[S_{j}^{z},H\right]=0$. Therefore, when we choose the eigenstates of the local spins $S_{j}^{z}$ as bases, we have $S_{j}^{z}|s_{j}\rangle=\frac{s_{j}}{2}|s_{j}\rangle$, where $s_{j}=\pm1$. For each configuration $\left\{  s_{j}\right\}$, we can express the Hamiltonian in terms of the following bilinear form,
\begin{align}
H_{s}(t)  & =-g\sum_{\langle ij\rangle\sigma}\left(  c_{i\sigma}^{\dagger}c_{j\sigma}+\text{H.c.}\right)  +\frac{J(t)}{4}\sum_{j\sigma}s_{j}\sigma^{z}c_{j\sigma}^{\dagger}c_{j\sigma}\nonumber\\
& =\Psi^{\dag}A\left(  s,t\right)  \Psi,
\end{align}
where $\Psi^{\dagger}=\left[c_{1\uparrow}^{\dagger},\cdots, c_{N\uparrow}^{\dagger},c_{1\downarrow}^{\dagger}, \cdots, c_{N\downarrow}^{\dagger}\right]$, $N=L \times L$ is the number of lattice sites, and $A(s,t)$ is its corresponding matrix representation.
In this case, the trace over the localized spin can be transformed into the summation over all possible configurations of $\left\{  s_{j}\right\}$,
\begin{equation}
\widetilde{G}\left(  R\right)  =\sum_{\left\{  s_{j}\right\}  }\operatorname{Tr}_{c}\left[  U^{\dag}\left(  \tau,0\right)  \operatorname{e}^{RH_{s}\left(  \tau\right)  }U\left(  \tau,0\right)  \operatorname{e}^{-\left(\beta+R\right)  H_{s}\left(  0\right)  }\right]  .
\end{equation}
Following the same computational paradigm as the trace evaluation in partition function, the trace of the conducting electron in $\widetilde{G}\left(  R\right)$ can likewise be computed through the standard framework of functional field integral \cite{Dong2019PRB},
\begin{equation}
\widetilde{G}\left(  R\right)  =\sum_{\left\{  s_{j}\right\}  }\det\left[I+B\right],
\end{equation}%
where the determinant comes from the Grassmann Gaussian integration over the fermionic coherent states, $I$ is the $N\times N$ identity matrix, and matrix $B$ is given by
\begin{align}
B  & =\operatorname{e}^{\operatorname*{i}\Delta tA\left(  s,t_{1,2}\right)}\cdots\operatorname{e}^{\operatorname*{i}\Delta tA\left(  s,t_{M,M+1}\right)}\operatorname{e}^{RA\left(  s,\tau\right)  }\nonumber\\
& \times\operatorname{e}^{-\operatorname*{i}\Delta tA\left(  s,t_{M,M+1}\right)  }\cdots\operatorname{e}^{-\operatorname*{i}\Delta tA\left(s,t_{1,2}\right)  }\operatorname{e}^{-\left( \beta+ R\right)  A\left(s,0\right)  }. \label{B_general_matrix}
\end{align}
In the path integral method, the time evolution period $\tau$ is segmented into $M$ minute intervals due to time-slicing. Obviously, $\Delta t = \tau/M$ directly affects the precision in numerical simulations.

Finally, introducing
\begin{equation}
Q\left(  \left\{  s_{j}\right\},R\right)  =-\ln\det\left[  I+B\right],
\end{equation}
we have $\widetilde{G}\left(  R\right)  = \sum_{\left\{  s_{j}\right\}}\operatorname{e}^{-Q \left(  \left\{  s_{j}\right\},R\right)  }$. The denominator of $G(R)$ in Eq. \eqref{gfgeneral} is the partition function at the initial time $t=0$, which can be denoted as $Z\left(  0\right)   =\widetilde{G}\left(  0\right)$. Now, the moment generating function becomes
\begin{equation}
G\left(  R\right)  =\frac{\sum_{\left\{  s_{j}\right\}  }\operatorname{e}^{-Q \left(  \left\{  s_{j}\right\}, R\right)  }}{\sum_{\left\{  s_{i}\right\}  }\operatorname{e}^{-Q \left(  \left\{  s_{j}\right\},0\right)  }}. \label{GR_final}
\end{equation}
There are two possible values of $s_{j}=\pm1$ at site $j$. The total configurations $\left\{ s_{j}\right\}$ amount to $2^{N}$. Direct summation in Eq. (\ref{GR_final}) becomes infeasible for large $N$. Instead, we can use the importance sampling to calculate Eq. \eqref{GR_final} by rewriting it as
\begin{equation}
G\left(  R\right)  =\sum_{\left\{  s_{j}\right\}  }P\left(  \left\{  s_{j}\right\} \right)\operatorname{e}^{-\left[  Q \left(  \left\{  s_{j}\right\},R\right)  -Q \left(  \left\{  s_{j}\right\},0\right)  \right]  }, \label{GR_Monte}
\end{equation}
where $P\left(  \left\{  s_{j}\right\} \right)$ is the probability of configuration $\left\{  s_{j}\right\}$,
\begin{equation}
P\left(  \left\{  s_{j}\right\} \right)=\frac{\operatorname{e}^{-Q \left(  \left\{  s_{j}\right\},0\right)  }}{\sum_{\left\{s_{i}\right\}  }\operatorname{e}^{-Q \left(  \left\{  s_{i}\right\},0\right)  }}. \label{Psj}
\end{equation}%
It is easy to see that $P\left( \left\{  s_{j}\right\} \right)$ satisfies the normalization condition $\sum_{\left\{  s_{j}\right\}  }P\left(  \left\{  s_{j}\right\}\right)  =1$. Equations~\eqref{GR_Monte} and \eqref{Psj} are essential to determine the moment generating function via the Monte Carlo method. This proposed approach can also be interpreted as a generalized determinant quantum Monte Carlo framework applicable to quantum thermodynamics in correlated electron systems, since the evaluation of transition probabilities inherently requires determinant calculations.

We use the usual Metropolis algorithm to calculate the probability distribution function $P\left(  \left\{  s_{j}\right\} \right)$.  The main Monte Carlo procedures involve: (1) Initiating with a random set of configuration $\left\{  s_{j}\right\}  =\left\{  s_{1},s_{2},\ldots,s_{N}\right\}$. (2) For each Monte Carlo sweep, we sequentially update one of the $N$ variables in $\left\{  s_{j}\right\}$, so that $s_{i}^{\prime}\rightarrow-s_{i}$ and $\left\{  s_{j}\right\}  ^{\prime}=\left\{  s_{1},\ldots,-s_{i},\ldots ,s_{N}\right\}$. Then we calculate $\Delta Q=Q\left(  \left\{  s_{j}\right\}  ^{\prime},0\right)  -Q\left(  \left\{  s_{j}\right\},0\right)$. (3) Following the Metropolis algorithm, the proposed configuration $s_{i}^{\prime}$ is accepted with probability $\min\left[  1,\operatorname{e}^{-\Delta Q}\right]$. This procedure is repeated sequentially for all $N$ variables in $\left\{  s_{j}\right\}$. (4) Repeat step (2) to thermalize the system. (5) After thermalization, 30000 Monte Carlo sweeps are performed for the Monte Carlo sampling. (6) Taking one sample (configuration) every $10$ sweeps to reduce self-correlations in the data and calculating the physical quantities by average over a total number
of $N_s=3000$ configurations. The generating function is now given by
\begin{equation}
G\left(  R\right)  \approx\frac{1}{N_{s}}\sum_{m=1}^{N_{s}}\operatorname{e}^{-\left[  Q\left(  \left\{  s_{j}\right\}_m ,R\right)  -Q\left(  \left\{  s_{j}\right\}_m ,0\right)  \right]  }.
\end{equation}

\begin{figure}[ptb]
\begin{center}
\includegraphics[width=8.6cm]{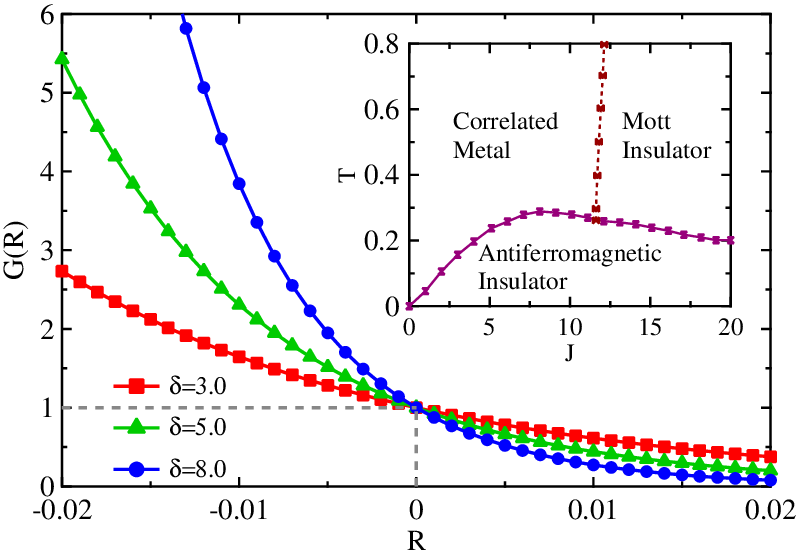}
\end{center}
\caption{The moment generating function $G(R)$ for three typical quench processes when $J$ changes from $J_0$ to $J_1=J_0+\delta$ at $T=0.2$. Here, the number of lattice sites is $N=16 \times 16$ and the initial Kondo coupling strength is $J_{0}=2.1$. The gray dashed lines indicate the normalization condition $G(0)=1$. The inset reproduces the equilibrium phase diagram of the Ising-Kondo lattice model \cite{Zhong2019PRB}. The solid line in the phase diagram at low temperature represents phase transition while the dashed lines represents crossover at high temperature.}
\label{fig1}
\end{figure}

The framework aforementioned remains valid for general nonequilibrium protocols $J(t)$ through the complete time evolution encoded in matrix $B$, as shown in Eq. \eqref{B_general_matrix}. In this work, we will focus on quench protocols for the sake of simplicity. When the control parameter $J(t)$ changes from $J_0$ at time $t=0^{-}$ to $J_1=J_0+\delta$ at time $t=0^{+}$, the time-dependent evolution $ U\left(  0^{+},0^{-}\right)$ becomes an unit operator and the matrix $B$ can be simplified as
\begin{equation}
B   =\operatorname{e}^{RA\left(  s,J_{1}\right)  }\operatorname{e}^{-\left(  \beta+R\right)  A\left(s,J_{0}\right)  }.
\end{equation}
Figure~{\ref{fig1}} illustrates the moment generating function for three distinct quench processes. As expected, all $G(R)$ curves pass smoothly through the point $G(0)=1$ and maintain $G(R)>0$ for all $R$.
The inset of Fig.~{\ref{fig1}} reproduces the phase diagram of the Ising-Kondo model at half-filling \cite{Zhong2019PRB}. The low-temperature metal-insulator transition is a phase transition, whereas at high temperatures it is a crossover.

After acquiring the moment generating function via the above Monte Carlo technique, evaluating the work statistics becomes straightforward. It is convenient to build the cumulant generating function $C\left(  R\right)$ from the moment generating function $C\left(  R\right)  =\ln G\left(  R\right)  $. Then a series expansion yields the cumulants $\kappa_{n}=\frac{d^{n}C\left(  R\right)  }{R^{n}}|_{R=0}$. According to statistical theory, the first cumulant is the mean work $\kappa_{1}=\left\langle W\right\rangle$. The second cumulant is the variance $\kappa_{2}=\sigma^{2}_{W}$, reflecting the fluctuations of work. The third cumulant is the third-order central moment $\kappa_{3}= \langle(W-\langle W\rangle)^{3}\rangle$, measuring the skewness and non-Gaussian fluctuations of work distributions. In addition to the cumulants, we can also obtain the free energy difference between $H(J_0)$ and $H(J_1)$ through the Jarzynski equality \cite{Jarzynski1997PRL,Jarzynski1997PRE}, $\Delta F=-\frac{1}{\beta}C\left(  -\beta\right)$. Then we can extract the mean irreversible work $ \langle W_{{\rm irr}}\rangle=\langle W\rangle-\Delta F$, which characterizes the irreversibility of the nonequilibrium process.

\begin{figure}[ptb]
\begin{center}
\includegraphics[width=8.6cm]{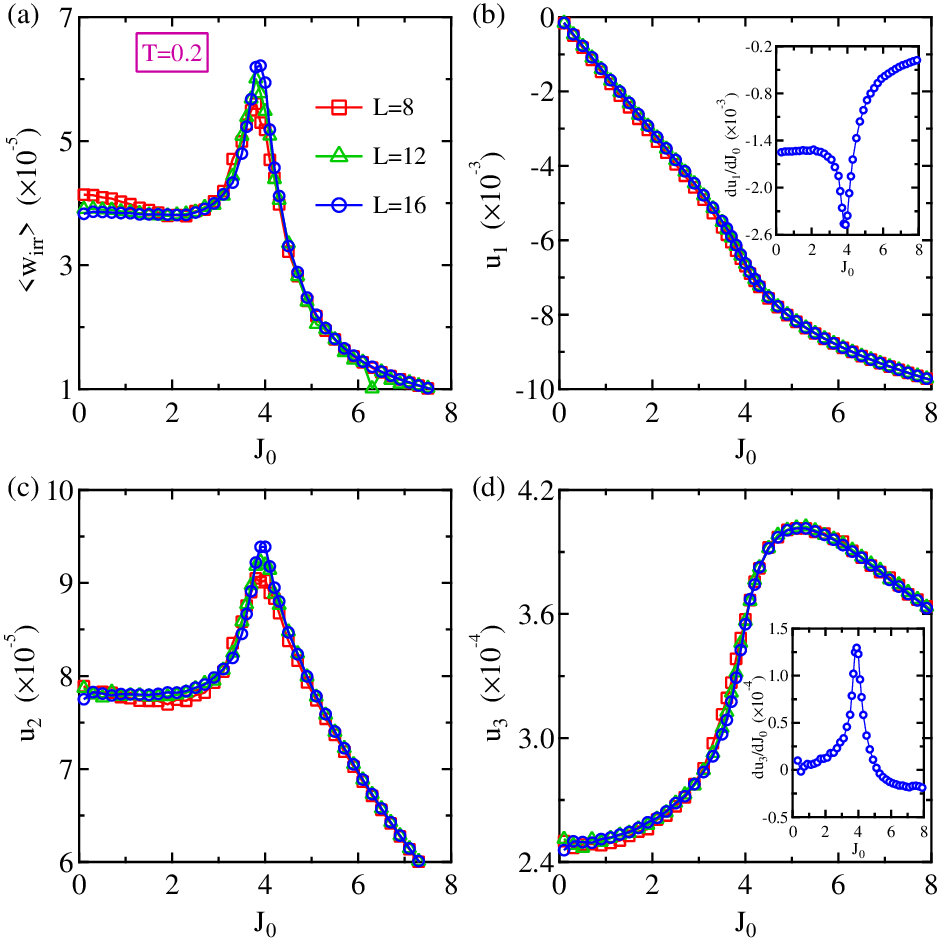}
\end{center}
\caption{(a) The mean irreversible work density $\langle w_{{\rm irr}}\rangle$, (b) scaled first cumulant (mean work density) $u_{1}$, (c) scaled second cumulant (variance) $u_{2}$, and (d) scaled third cumulant $u_{3}$ of quantum work as a function of $J_0$ for fixed $\delta=0.05$ at low temperature $T=0.2$. The insets in (b) and (d) are the derivatives of $u_{1}$ and $u_{3}$, i.e., $du_{1}/dJ_{0}$ and $du_{3}/dJ_{0}$ for $L=16$, respectively.}
\label{fig2}
\end{figure}

Due to the extensive properties, we will focus on the mean irreversible work density $\langle w_{{\rm irr}}\rangle = \langle W_{{\rm irr}}\rangle/N$ and the scaled cumulants $u_{n}=\kappa_{n}/N$. Figure~{\ref{fig2}} presents the average irreversible work density and the first three scaled cumulants at low temperatures with a quench amplitude of $\delta =0.05$. The mean irreversible work density and the scaled second cumulant (variance) exhibit a clear cusp at the phase transition point between the correlated metal phase and the antiferromagnetic insulator phase. The scaled first cumulant (mean work density) $u_{1}$ and the third cumulant $u_{3}$ show the steepest slope at the phase transition point, as shown in the inset of Fig.~{\ref{fig2}}(b) and {\ref{fig2}}(d).

\begin{figure}[ptb]
\begin{center}
\includegraphics[width=8.6cm]{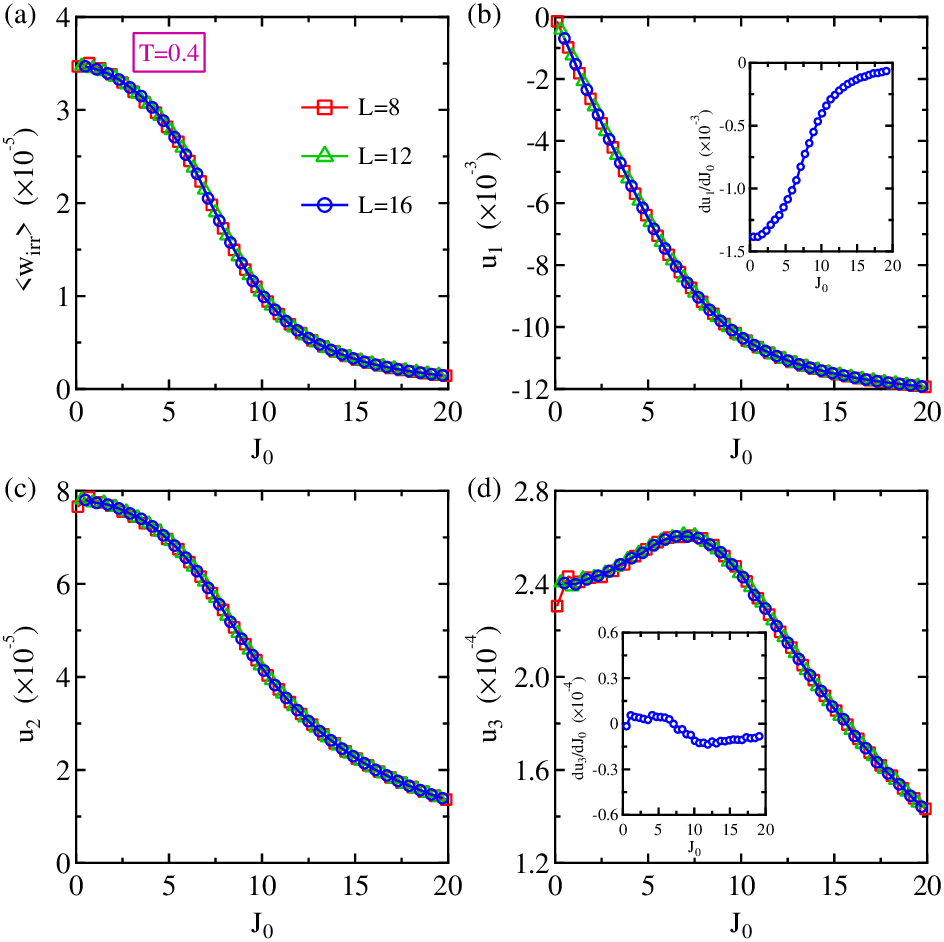}
\end{center}
\caption{(a) The mean irreversible work density $\langle w_{{\rm irr}}\rangle$, (b) scaled first cumulant (mean work density) $u_{1}$, (c) scaled second cumulant (variance) $u_{2}$, and (d) scaled third cumulant $u_{3}$ of quantum work as a function of $J_0$ for fixed $\delta=0.05$ at high temperature $T=0.4$. For comparison, the derivatives of $u_{1}$ and $u_{3}$ for $L=16$ are also presented in the insets of (b) and (d).}
\label{fig3}
\end{figure}

To investigate whether these characteristic features observed at low temperatures persist at high temperatures, we depict the mean irreversible work density and the first three scaled cumulants at an elevated temperature $T=0.4$ where the metal-insulator transition evolves into a smooth crossover. As shown in Fig.~\ref{fig3}, all the singularities observed at low temperatures vanish at high temperatures. This sharp dichotomy indicates that quantum work serves as a sensitive thermodynamic tool for identifying metal-insulator phase transition, a prototypical phenomenon in correlated electron systems which has been scarcely investigated within the framework of quantum thermodynamics.

In summary, we have developed a Monte Carlo framework to investigate the statistical properties of quantum work in strongly correlated electron systems, with the Ising-Kondo lattice model serving as a prototypical platform. By computing the moment generating function of quantum work under nonequilibrium quench protocols, we systematically examined critical statistical measures, such as the mean irreversible work density, mean work density, variance, and third central moment across diverse quench processes. Our findings demonstrate pronounced singularities in these quantities at the metal-insulator phase transition point under low-temperature conditions, which are entirely absent at elevated temperatures as the transition shifts to a smooth crossover. This striking dichotomy emphasizes the efficacy of quantum work as a thermodynamic probe for detecting quantum phase transitions. The developed Monte Carlo methodology, grounded in determinant quantum Monte Carlo techniques, effectively addresses challenges inherent to strongly correlated systems and establishes a versatile framework for exploring nonequilibrium quantum thermodynamics. By linking quantum work statistics to critical phenomena in correlated materials, this work paves the way for investigating dynamical phase transitions and irreversibility in complex quantum systems. Our results elucidate the intricate interplay between quantum coherence, thermal fluctuations, and nonequilibrium dynamics, offering insights applicable to a wide range of strongly correlated systems.

This work is supported by the National Natural Science Foundation of China Grants No. 12204075, 12474140 and No. 12347101; the China Postdoctoral Science Foundation Grants No. 2023M730420; the fellowship of the Chongqing Postdoctoral Program for Innovative Talents Grant No. CQBX202222; the Natural Science Foundation of Chongqing Grant No. CSTB2023NSCQ-MSX0953, No. CSTB2023NSCQ-MSX0048, and No. CSTB2024YCJH-KYXM0064.


\begin{thebibliography}{99}

\bibitem {Binder2018} F. Binder, L. A. Correa, C. Gogolin, J. Anders, and G. Adesso, \textit{Thermodynamics in the Quantum Regime}, (Springer, Cham, Switzerland, 2018).
\bibitem {Deffner2019} S. Deffner and S. Campbell, \textit{Quantum Thermodynamics: An introduction to the thermodynamics of quantum information}, (Morgan $\&$ Claypool, San Rafael, CA, 2019).
\bibitem {Alicki2023PRD} R. Alicki, G. Barenboim, and A. Jenkins, Quantum thermodynamics of de Sitter space, Phys. Rev. D \textbf{108}, 123530 (2023).
\bibitem {Gatto2024PRA} S. Gatto, A. Colla, H.-P. Breuer, and M. Thoss, Quantum thermodynamics of the spin-boson model using the principle of minimal dissipation, Phys. Rev. A \textbf{110}, 032210 (2024).
\bibitem {WeiWu2024PRL} W. Wu and J.-H. An, Generalized Quantum Fluctuation Theorem for Energy Exchange, Phys. Rev. Lett. \textbf{133}, 050401 (2024).
\bibitem {Kosloff2013Entropy} R. Kosloff, Quantum Thermodynamics: A Dynamical Viewpoint, Entropy, \textbf{15}, 2100 (2013).
\bibitem {Vinjanampathy2016ContempPhys} S. Vinjanampathy and J. Anders, Quantum Thermodynamics, Contemp. Phys. \textbf{57}, 545 (2016).
\bibitem {Millen2016NJP} J. Millen and A. Xuereb, Perspective on quantum thermodynamics, New J. Phys. \textbf{18}, 011002 (2016).
\bibitem {Dann2023NJP} R. Dann and R. Kosloff, Unification of the first law of quantum thermodynamics, New J. Phys. \textbf{25}, 043019 (2023).

\bibitem {Ortega2019PRL} A. Ortega, E. Mckay, \'A. M. Alhambra, and E. M.-Mart\'{\i}nez, Work Distributions on Quantum Fields, Phys. Rev. Lett. \textbf{122}, 240604 (2019).
\bibitem {Beyer2020PRR} K. Beyer, K. Luoma, and W. T. Strunz, Work as an external quantum observable and an operational quantum work fluctuation theorem, Phys. Rev. Res. \textbf{2}, 033508 (2020).
\bibitem {Pourhassan2021JHEP} B. Pourhassan, S. S. Wani, S. Soroushfarc and M. Faizal, Quantum work and information geometry of a quantum Myers-Perry black hole, J. High Energ. Phys. \textbf{2021}, 27 (2021).
\bibitem {Gardas2016SciRep} B. Gardas, S. Deffner, and A. Saxena, Non-hermitian quantum thermodynamics, Sci. Rep. \textbf{6}, 23408 (2016).
\bibitem {Wei2018PRE} B.-B. Wei, Quantum work relations and response theory in parity-time-symmetric quantum systems, Phys. Rev. E \textbf{97}, 012114 (2018).
\bibitem {Zhou2021PRE} Z.-Y. Zhou, Z.-L. Xiang, J. Q. You, and F. Nori, Work statistics in non-Hermitian evolutions with Hermitian endpoints, Phys. Rev. E \textbf{104}, 034107 (2021).
\bibitem {Erdamar2024PRR} S. Erdamar, M. Abbasi, B. Ha, W. Chen, J. Muldoon, Y. Joglekar, and K. W. Murch, Constraining work fluctuations of non-Hermitian dynamics across the exceptional point of a superconducting qubit, Phys. Rev. Res. \textbf{6}, L022013 (2024).
\bibitem {Lacerda2023PRB} A. M. Lacerda, A. Purkayastha, M. Kewming, G. T. Landi, and J. Goold, Quantum thermodynamics with fast driving and strong coupling via the mesoscopic leads approach, Phys. Rev. B \textbf{107}, 195117 (2023).
\bibitem {Lacerda2024PRE} A. M. Lacerda, M. J. Kewming, M. Brenes, C. Jackson, S. R. Clark, M. T. Mitchison, and J. Goold, Entropy production in the mesoscopic-leads formulation of quantum thermodynamics, Phys. Rev. E \textbf{110}, 014125 (2024).
\bibitem {Binder2023PRL} D. \v{S}afr\'{a}nek, D. Rosa, and F. C. Binder, Work Extraction from Unknown Quantum Sources, Phys. Rev. Lett. \textbf{130}, 210401 (2023).
\bibitem {Gherardini2024PRXQuant} S. Gherardini and G. D. Chiara, Quasiprobabilities in Quantum Thermodynamics and Many-Body Systems, PRX Quantum \textbf{5}, 030201 (2024).
\bibitem {Davoudi2024PRL} Z. Davoudi, C. Jarzynski, N. Mueller, G. Oruganti, C. Powers, and N. Y. Halpern, Quantum Thermodynamics of Nonequilibrium Processes in Lattice Gauge Theories, Phys. Rev. Lett. \textbf{133}, 250402 (2024).


\bibitem {Lobejko2017PRE} M. Lobejko, J. Luczka, and P. Talkner, Work distributions for random sudden quantum quenches, Phys. Rev. E \textbf{95}, 052137 (2017).
\bibitem {Francica2023PRB} G. Francica and L. D. Anna, Quasiprobability distribution of work in the quantum Ising model, Phys. Rev. B \textbf{108}, 014106 (2023).
\bibitem {Santini2023PRB} A. Santini, A. Solfanelli, S. Gherardini, and M. Collura, Work statistics, quantum signatures, and enhanced work extraction in quadratic fermionic models, Phys. Rev. B \textbf{108}, 104308 (2023).
\bibitem {Kiely2023PRR} A. Kiely, E. O'Connor, T. Fogarty, G. T. Landi, and S. Campbell, Entropy of the quantum work distribution, Phys. Rev. Res. \textbf{5}, L022010 (2023).
\bibitem {Pal2023PRB} K. Pal, K. Pal, A. Gill, and T. Sarkar, Time evolution of spread complexity and statistics of work done in quantum quenches, Phys. Rev. B \textbf{108}, 104311 (2023).
\bibitem {Zawadzki2023PRA} K. Zawadzki, A. Kiely, G. T. Landi, and S. Campbell, Non-Gaussian work statistics at finite-time driving, Phys. Rev. A \textbf{107}, 012209 (2023).

\bibitem {Jarzynski1997PRL} C. Jarzynski, Nonequilibrium Equality for Free Energy Differences, Phys. Rev. Lett. \textbf{78}, 2690 (1997).
\bibitem {Jarzynski1997PRE} C. Jarzynski, Equilibrium free-energy differences from nonequilibrium measurements: A master-equation approach, Phys. Rev. E \textbf{56}, 5018 (1997).
\bibitem {Silva2008PRL} A. Silva, Statistics of the Work Done on a Quantum Critical System by Quenching a Control Parameter, Phys. Rev. Lett. \textbf{101}, 120603 (2008).
\bibitem {Fei2020PRL} Z. Fei, N. Freitas, V. Cavina, H. T. Quan, and M. Esposito, Work Statistics across a Quantum Phase Transition, Phys. Rev. Lett. \textbf{124}, 170603 (2020).
\bibitem {Wu2025PRL} Y.-X. Wu, J.-Fu Chen, and H. T. Quan, Ergodicity Breaking and Scaling Relations for Finite-Time First-Order Phase Transition, Phys. Rev. Lett. \textbf{134}, 177101 (2025).
\bibitem {Solfanelli2025PRL} A. Solfanelli and N. Defenu, Universal Work Statistics in Long-Range Interacting Quantum Systems, Phys. Rev. Lett. \textbf{134}, 030402 (2025).
\bibitem {Heyl2013PRL} M. Heyl, A. Polkovnikov, and S. Kehrein, Dynamical Quantum Phase Transitions in the Transverse-Field Ising Model, Phys. Rev. Lett. \textbf{110}, 135704 (2013).
\bibitem {Campisi2017PRE} M. Campisi and J. Goold, Thermodynamics of quantum information scrambling, Phys. Rev. E \textbf{95}, 062127 (2017).
\bibitem {Chenu2018SciRep} A. Chenu, I. L. Egusquiza, J. M.-Vilaplana, and A. D. Campo, Quantum work statistics, Loschmidt echo and information scrambling, Sci. Rep. \textbf{8}, 12634 (2018).


\bibitem {FeiLiu2012PRER} F. Liu, Derivation of quantum work equalities using a quantum Feynman-Kac formula, Phys. Rev. E \textbf{86}, 010103(R) (2012).
\bibitem {Funo2018PRL} K. Funo and H. T. Quan, Path Integral Approach to Quantum Thermodynamics, Phys. Rev. Lett. \textbf{121}, 040602 (2018).
\bibitem {Dong2019PRB} J.-J. Dong and Y.-F. Yang, Functional field integral approach to quantum work, Phys. Rev. B \textbf{100}, 035124 (2019).
\bibitem {Quan2020PRE} T. Qiu, Z. Fei, R. Pan, and H. T. Quan, Path-integral approach to the calculation of the characteristic function of work, Phys. Rev. E \textbf{101}, 032111 (2020).
\bibitem {FeiLiu2019PRE} Y. Qian and F. Liu, Computing characteristic functions of quantum work in phase space, Phys. Rev. E \textbf{100}, 062119 (2019).
\bibitem {Quan2019PRR} Z. Fei and H. T. Quan, Group-theoretical approach to the calculation of quantum work distribution, Phys. Rev. Res. \textbf{1}, 033175 (2019).
\bibitem {Quan2020PRL} Z. Fei and H. T. Quan, Nonequilibrium Green's Function's Approach to the Calculation of Work Statistics, Phys. Rev. Lett. \textbf{124}, 240603 (2020).
\bibitem {Grabarits2022SciRep} A. Grabarits, M. Kormos, I. Lovas, and G. Zar\'{a}nd, Classical theory of universal quantum work distribution in chaotic and disordered non-interacting Fermi systems, Sci. Rep. \textbf{12}, 15017 (2022).
\bibitem {Quan2022PRR} J. Gu, F. Zhang, and H. T. Quan, Tensor-network approach to work statistics for one-dimensional quantum lattice systems, Phys. Rev. Res. \textbf{4}, 033193 (2022).
\bibitem {FengLiLin2024PRR} F.-L. Lin and C.-Y. Huang, Work statistics for quantum spin chains: Characterizing quantum phase transitions, benchmarking time evolution, and examining passivity of quantum states, Phys. Rev. Res. \textbf{6}, 023169 (2024).


\bibitem {Gubernatis2016} J. E. Gubernatis, N. Kawashima, and P. Werner, \textit{Quantum Monte Carlo Methods: Algorithms for Lattice Models}, (Cambridge University Press, Cambridge, 2016).
\bibitem {Farkasovsky2010PRB} P. Farka\v{s}ovsk\'{y}, H. \v{C}en\v{c}arikov\'{a}, and S. Mata\v{s}, Numerical study of magnetization processes in rare-earth tetraborides, Phys. Rev. B \textbf{82}, 054409 (2010).
\bibitem {Kato2010PRL} Y. Kato, I. Martin, and C. D. Batista, Stability of the Spontaneous Quantum Hall State in the Triangular Kondo-Lattice Model, Phys. Rev. Lett. \textbf{105}, 266405 (2010).
\bibitem {Ishizuka2012PRL} H. Ishizuka and Y. Motome, Partial Disorder in an Ising-Spin Kondo Lattice Model on a Triangular Lattice, Phys. Rev. Lett. \textbf{108}, 257205 (2012).
\bibitem {Maska2020PRB} M. M. Maska and N. Trivedi, Temperature-driven BCS-BEC crossover and Cooper-paired metallic phase in coupled boson-fermion systems, Phys. Rev. B \textbf{102}, 144506 (2020).
\bibitem {Dong2021PRBL} J.-J. Dong, D. Huang, and Y.-F. Yang, Mutual information, quantum phase transition, and phase coherence in Kondo systems, Phys. Rev. B \textbf{104}, L081115 (2021).
\bibitem {Dong2022PRBL} J.-J. Dong and Y.-F. Yang, Development of long-range phase coherence on the Kondo lattice, Phys. Rev. B \textbf{106}, L161114 (2022).
\bibitem {Qin2023PRB} Q. Qin, J.-J. Dong, Y. Sheng, D. Huang, and Y.-F. Yang, Superconducting fluctuations and charge-4e plaquette state at strong coupling, Phys. Rev. B \textbf{108}, 054506 (2023).


\bibitem {Sikkema1996PRB} A. E. Sikkema, W. J. L. Buyers, I. Affleck, and J. Gan, Ising-Kondo lattice with transverse field: A possible f-moment Hamiltonian for URu$_2$Si$_2$, Phys. Rev. B \textbf{54}, 9322 (1996).
\bibitem {Zhong2019PRB} W.-W. Yang, J. Zhao, H.-G. Luo, and Y. Zhong, Exactly solvable Kondo lattice model in the anisotropic limit, Phys. Rev. B \textbf{100}, 045148 (2019).
\bibitem {Zhong2021PRB} W.-W. Yang, Y.-X. Li, Y. Zhong, and H.-G. Luo, Doping a Mott insulator in an Ising-Kondo lattice: Strange metal and Mott criticality
, Phys. Rev. B \textbf{104}, 165146 (2021).
\bibitem {Zhou2024PRB} X. Zhou, J. Fan, and S. Jia, Magnetic order and strongly correlated effects in the one-dimensional Ising-Kondo lattice, Phys. Rev. B \textbf{109}, 195112 (2024).





\end{thebibliography}
\end{document}